\documentclass{ws-mpla}

\def\bc{\begin{center}}
\def\ec{\end{center}}
\def\be{\begin{equation}}
\def\ee{\end{equation}}
\def\trm{\textrm}

\def\GE{G_{\trm{E}}}
\def\EMSR{\langle r^{2} \rangle _{\trm{E}}}

\begin{document}

\markboth{T. Sekihara, T. Hyodo, D. Jido}
{Electric Mean Squared Radii of $\Lambda (1405)$ in Chiral Dynamics}

\catchline{}{}{}{}{}

\title{ELECTRIC MEAN SQUARED RADII OF $\Lambda (1405)$ 
IN CHIRAL DYNAMICS}

\author{\footnotesize T. SEKIHARA
}

\address{Department of Physics, Kyoto University, 606-8502
Kyoto, Japan\\
sekihara@ruby.scphys.kyoto-u.ac.jp
}

\author{T. HYODO}

\address{Physik-Department, Technische Universit\"{a}t M\"{u}nchen,
 D-85747 Garching, Germany, and\\
Yukawa Institute for Theoretical Physics, Kyoto University, 
Kyoto 606--8502, Japan 
}

\author{D. JIDO}

\address{Yukawa Institute for Theoretical Physics, 
Kyoto University, 606-8502
Kyoto, Japan
}

\maketitle

\pub{Received (Day Month Year)}{Revised (Day Month Year)}

\begin{abstract}
The electric mean squared radii $\EMSR$ of $\Lambda (1405)$ 
are calculated in the chiral unitary model. We describe the 
$\Lambda(1405)$ as a dynamically generated resonance 
fully in the octet meson and octet baryon scattering. 
We also consider ``$\Lambda(1405)$'' as a bound state of 
$\bar{K} N$. For the later ``$\Lambda (1405)$,'' we obtain 
negative and larger absolute value of 
electric mean squared radius than that of ordinary baryons, which
implies that $\Lambda (1405)$ have structure of widely spread 
$K^{-}$ around $p$.

\end{abstract}

\ccode{PACS Nos.: 13.75.Jz, 14.20.-c, 11.30.Rd}

\section{Introduction}	
The $\Lambda(1405)$ has been considered as a quasi-bound state 
of anti-kaon and nucleon ($\bar{K} N$) system for a long 
time.\cite{Dalitz} 
Recent theoretical investigations have also suggested that 
the $\Lambda(1405)$ is well described as a dynamically generated
resonance of the meson-baryon scattering based on chiral dynamics 
in coupled channel approach,\cite{Kaiser,Oset,Oller,Lutz,Jido} which
is called chiral unitary model (ChUM). In ChUM, the scattering amplitude
is obtained by an algebraic equation:
\be
T_{ij}(\sqrt{s}) = V_{ij}(\sqrt{s}) + \sum _{k} V_{ik} (\sqrt{s}) 
G_{k} (\sqrt{s}) T_{kj} (\sqrt{s}) ,
\label{eq:BSEq}
\ee
with an $s$-wave interaction kernel $V$ given by the 
lowest order chiral perturbation
theory, that is the Weinberg-Tomozawa interaction, and the meson-baryon
loop function $G$, which are functions of the 
center-of-mass energy 
of the meson-baryon system, $\sqrt{s}$, in matrix forms of 
the meson-baryon channels of strangeness $-1$ and charge $0$. 
ChUM has reproduced well the scattering cross 
sections of $K^{-}p$ to
various channels and the mass spectrum of the $\Lambda(1405)$ 
resonance below the $\bar{K} N$ threshold, giving two states for 
the $\Lambda(1405)$ as poles of the scattering amplitudes
in the complex energy plane, 
($z_{1}=1390-66i \, \trm{MeV}$) and 
($z_{2}=1426-17i \, \trm{MeV}$).\cite{Oller,Jido} 
The $\Lambda(1405)$ plays an important role for the system of 
kaons and nucleons, such as kanoic nuclei.\cite{HW} 

Since the $\Lambda(1405)$ is sitting below the $\bar{K} N$ 
threshold and it is difficult to obtain precise and clean information 
of the $\Lambda(1405)$
in the present experiments, any theoretical investigations are welcome. 
If the $\Lambda(1405)$ is described by the quasi-bound state of the
meson-baryon system with small binding energy, 
one expects that the $\Lambda(1405)$ has 
a larger size than typical ground state baryons dominated by a 
genuine quark state. If this is the case, the form factor of the 
$\Lambda(1405)$ falls off more rapidly 
than that of the nucleon and the
production cross section of the $\Lambda(1405)$ 
has large energy dependence.  
In this work, we estimate the electric mean squared radii of 
the $\Lambda(1405)$ based on the chiral unitary model. 

\section{Form Factors and Mean Squared Radii of Excited Baryons}

In this section, we discuss the formulation to evaluate the electric 
form factors and the charge radii of the $\Lambda(1405)$. 
First of all, let us define the electric form factor of an excited baryon 
$|H^{\ast}\rangle$, as a matrix element of zeroth component of 
the electromagnetic current 
$J_{\trm{EM}}^{\mu}$ in the Breit frame:\cite{Jido2} 
\be
\left \langle H^{\ast} \left | J_{\trm{EM}}^{0} \right | H^{\ast}
\right \rangle \equiv \GE (Q^{2}) ,
\ee
with the virtual photon momentum $q^{\mu}$ and $Q^{2}=-q^{2}$.
Using this definition, the form factor of the excited baryon can be
expressed as a residue of the second rank pole of the 
$MB \gamma ^{\ast} \rightarrow MB$ amplitude 
$T^{\mu}_{\gamma ij}$ with $\mu=0$: 
\begin{align}
- i T_{\gamma ij}^{0} & \simeq 
(- i g_{i})\frac{i}{\sqrt{s} - z_{\trm{H}}} 
\left \langle H^{\ast} \left | i J_{\trm{EM}}^{0} \right | H^{\ast}
\right \rangle 
\frac{i}{\sqrt{s} - z_{\trm{H}}} (- i g_{j}) \nonumber \\ 
&= (- i g_{i})\frac{i}{\sqrt{s} - z_{\trm{H}}} 
\big[ i \GE (Q^{2}) \big]
\frac{i}{\sqrt{s} - z_{\trm{H}}} (- i g_{j}) , 
\label{eq:Tgamma}
\end{align}
where $z_{\trm{H}}$ is the pole position and 
$g_{i}$ is the coupling strength of the excited baryon 
to the meson-baryon state. When an excited baryon is 
dynamically generated in ChUM, the scattering amplitude 
close to the pole of the excited baryon can be expressed by, 
\be
- i T_{ij} (\sqrt{s}) \simeq (- i g_{i})
\frac{i}{\sqrt{s} - z_{\trm{H}}}(- i g_{j}) .
\label{eq:T}
\ee
Combining Eqs.~(\ref{eq:Tgamma}) and (\ref{eq:T}), 
the form factor can be evaluated as a residue of the pole at 
$\sqrt{s} = z_{\trm{H}}$ as discussed in Ref.~\refcite{Jido2}:
\be
\GE (Q^{2}) = 
\trm{Res} \left[ - \frac{T_{\gamma ij}^{0}}{T_{ij}} \right] 
= \trm{Res} \left[ \frac{\GE (Q^{2})}
{\sqrt{s} - z_{\trm{H}}} \right] .
\ee
Then we can define electric mean squared radii $\EMSR$ 
as the usual form;
\be
\EMSR \equiv - 6 \left . 
\frac{d \GE}{d Q^{2}} \right | _{Q^{2}=0}.
\ee

\begin{figure}[t]
 \bc
 \begin{tabular}{ccc}
    \includegraphics[scale=0.115]{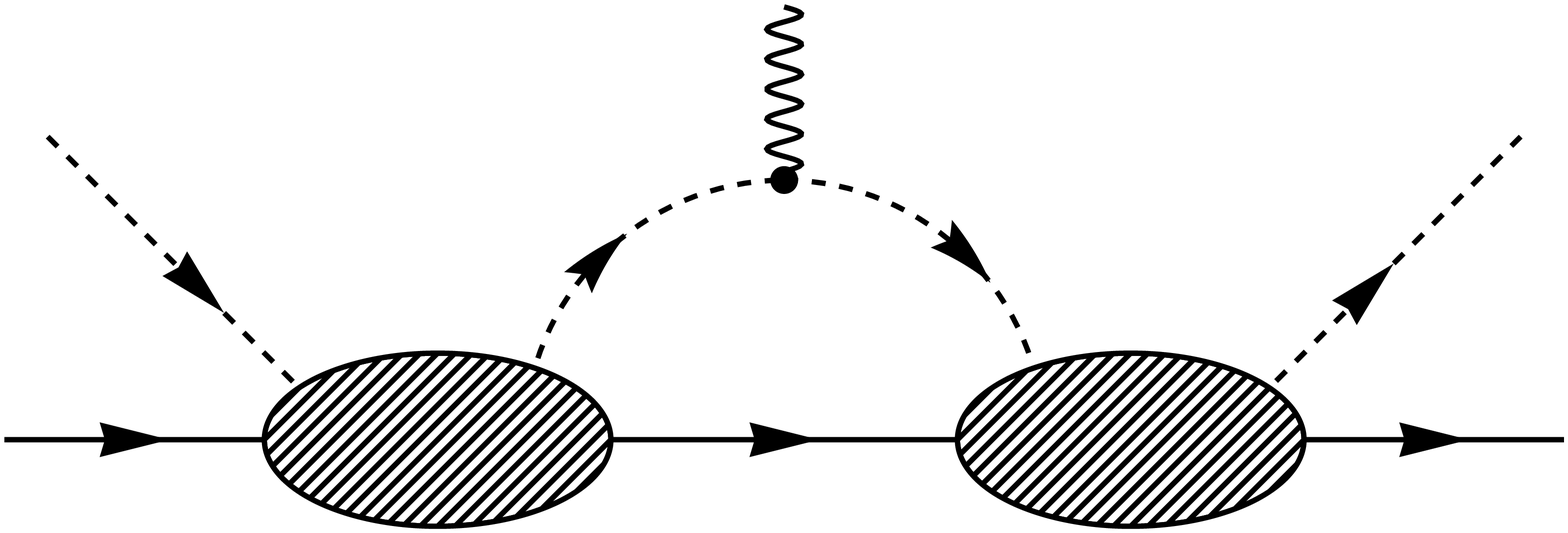} &
    \includegraphics[scale=0.115]{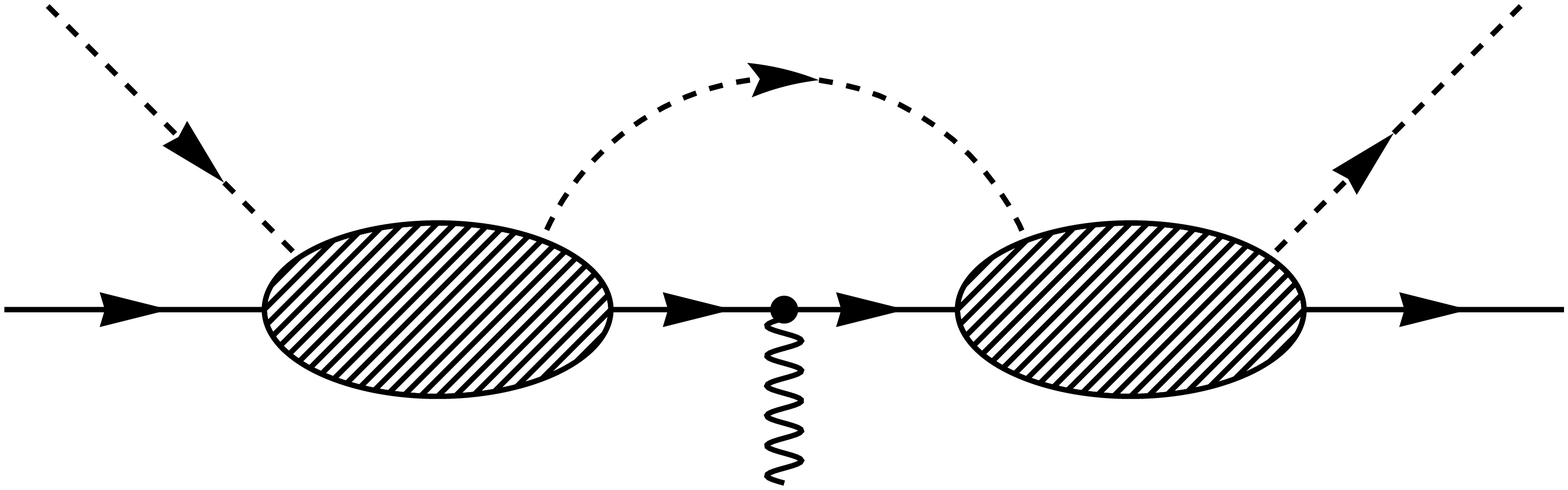} &
    \includegraphics[scale=0.115]{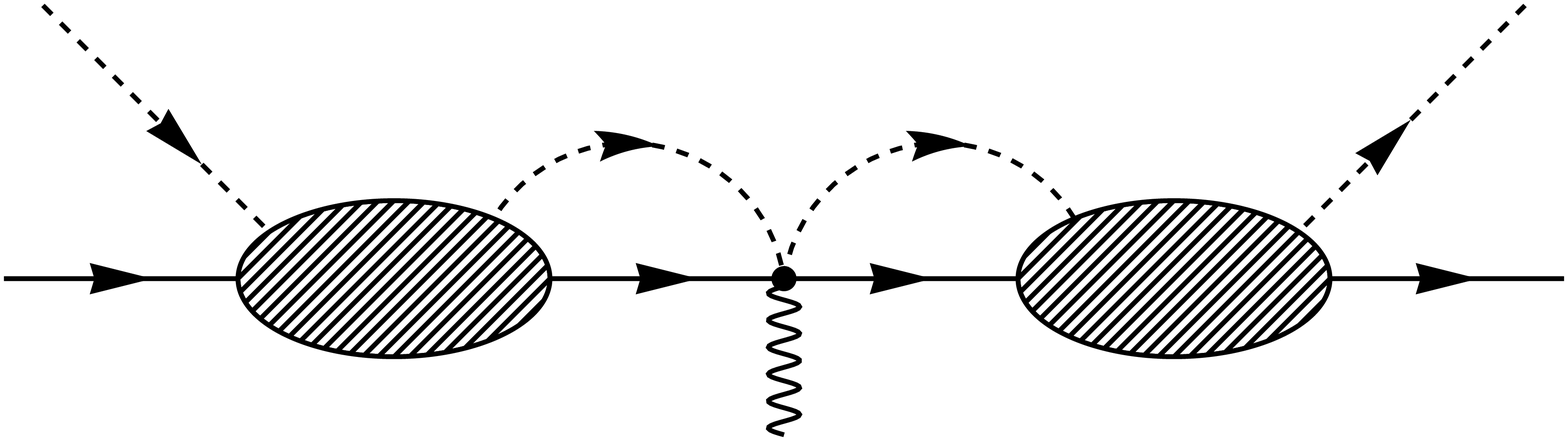} \\
    $- i T_{\gamma 1}^{\mu}$ &
    $-iT_{\gamma 2}^{\mu}$ &
    $-iT_{\gamma 3}^{\mu}$
 \end{tabular}
 \ec
 \caption{Diagrams for the form factor of the $\Lambda(1405)$. 
The shaded ellipses represent the meson-baryon scattering amplitudes.
 \label{diagram}}
\end{figure}

In our approach, the amplitude $T_{\gamma ij}^{\mu}$ is calculated 
under the assumption that the photon couples to the $\Lambda(1405)$ 
through the photon couplings to the constituent meson and baryon
of the $\Lambda(1405)$. 
The calculation of the photon coupling should be performed in a gauge
invariant way, since the gauge invariance guarantees to give 
the correct electric charge to the excited baryons, 
$\GE(Q^{2}=0)=Q_{\trm{H}}$.
Following the method of the gauge invariant calculation 
proposed by Ref.~\refcite{Borasoy}, 
we find the relevant diagrams for our purpose, which have the second 
rank poles, as shown in Fig.~\ref{diagram}. 
The shaded ellipses represent the meson-baryon scattering
amplitude $T_{ij}$ obtained by Eq.~(\ref{eq:BSEq}) 
and the photon couplings to the 
mesons and baryons are given by gauging the kinetic terms 
and the effective interaction. Calculating these three 
diagrams and summing up them,
we obtain, 
\be
T_{\gamma ij}^{\mu} \equiv T_{\gamma 1 ij}^{\mu} + 
T_{\gamma 2 ij}^{\mu} + T_{\gamma 3 ij}^{\mu} .
\ee
The detailed calculation is given in Ref.~\refcite{Sekihara}.

\section{Numerical Results}

\begin{table}[b]
\tbl{Electric mean squared radii $\EMSR$ of $\Lambda (1405)$. 
Physical $\Lambda(1405)$ is accompanied by the poles $z_1$ and $z_2$.
``$\Lambda(1405)$'' is described as a $\bar{K}N$ bound state.
\label{result} }
{\begin{tabular}{@{}cccc@{}} \toprule
State & Pole position [MeV] & Strongly couple to & 
$\EMSR$ [$\trm{fm}^{2}$] \\
   \colrule
   $z_1$
   & 1390 $-$ 66 $i$
   & $\pi \Sigma$
   & $\phantom{-} (1.8 - 0.2 i ) \times 10^{-2}$ \\
   $z_2$
   & 1426 $-$ 17 $i$
   & $\bar{K} N$ 
   & $-0.131 + 0.303 i$ \\
   ``$\Lambda(1405)$'' 
   & 1429\phantom{ $-$ 17 $i$} 
   & $K^{-} p$, $\bar{K}^{0} n$ & 
   $-2.193$ \\
   \botrule
\end{tabular}}
\end{table}

We show our result for the electric mean squared radii of the $\Lambda(1405)$
in Table~\ref{result}. As mentioned before, we have two 
$\Lambda(1405)$ states; 
the lower state $z_{1}$ strongly couples to the 
$\pi\Sigma$ state, while the higher 
state $z_{2}$ dominantly couples to the $\bar{K} N$ state.\cite{Jido} 
Since the $\pi^{+}\Sigma^{-}$ and $\pi^{-}\Sigma^{+}$ 
contribute to the isospin $0$ state almost equally,
the electric mean squared radius of the lower $\Lambda (1405)$ 
state $z_{1}$ is 
suppressed. Since we have evaluated the electric radii of the 
$\Lambda(1405)$ at the resonant point, they are free from the 
non-resonant 
meson-baryon background and are pure resonance properties. But 
the values get complex numbers, because we calculate $\EMSR$ 
of baryon resonant states with complex energies. 
Experimentally observables may be the ratio of the amplitudes 
$T_{\gamma ij}^{0}/T_{ij}$ in the real energies,
which however includes the contributions from the non-resonant 
meson-baryon scattering states.\cite{Jido2} 

For the resonant states with small decay widths, it is possible to 
interpret the electric mean squared radii as the sizes, since the radii are 
close to real numbers. For this purpose, we analyze a 
bound state obtained by switching off the other channels than 
$\bar{K} N$.\cite{HW} 
This bound state is considered to be the origin of the higher state 
of the physical $\Lambda(1405)$, $z_{2}$. We 
thus estimate the size of a meson-baryon resonance by analyzing this 
``$\Lambda(1405)$'' described as a $\bar{K}N$ bound state. 
When we solve 
Eq.~\eqref{eq:BSEq} with $K^-p$ and $\bar{K}^0n$ channels, 
a bound state is 
found at 1429 MeV, and the result of the $\EMSR$ is shown in 
Table~\ref{result}. The obtained $\EMSR$ has a negative sign, which is 
reflected by the charge distribution that the $K^{-}$ is surrounding around 
the $p$. The absolute value of our result, 
$\sqrt{| \EMSR |} \simeq 1.48 \, \trm{fm}$, suggests that the 
$\Lambda(1405)$ has a larger size than that of the ordinary 
baryons such as proton. In the end, we note that the mean squared 
radii of $\Lambda (1405)$ decrease as the binding energies 
increase in our model, as seen in Table~\ref{result}. 
The detailed discussions are given in Ref.~\refcite{Sekihara}.

\section{Summary}

We have calculated the electric mean squared radii of $\Lambda (1405)$ based on the 
meson-baryon picture of the $\Lambda(1405)$ in the chiral unitary model. The 
evaluation of the electric mean squared radii have been performed in two 
ways: In the first approach, with full coupled channels for the 
$\Lambda(1405)$, the mean squared radii in complex numbers are obtained at 
the poles for the physical resonance. In the second, we describe 
``$\Lambda(1405)$'' as a bound state of $\bar{K} N$ by neglecting all the
coupling of $\bar{K} N$ to the other channels, in order to estimate the size of
the resonance. Our result suggests that the $\Lambda (1405)$ have structure of widely 
spread $K^{-}$ around $p$ and the size of the $\Lambda(1405)$ is 
larger than that of the ordinary baryons.

\end{document}